# A Numerical Study on Spray Characteristics at Start of Injection for Gasoline Direct Injection

Zongyu Yue[1*], Michele Battistoni[2], Sibendu Som[1]

[1]Argonne National Laboratory, USA

[2]University of Perugia, Italy

**Abstract**

This paper presents a CFD study of Engine Combustion Network (ECN) Spray G, focusing on the transient characteristics of spray at start of injection. The Large Eddy Simulation (LES) coupled with Volume of Fluid (VOF) method is used to model the turbulent two-phase flow. A moving needle boundary condition is applied to capture the internal flow condition accurately. The injector geometry was measured with micron-level resolution using full spectrum x-ray tomographic imaging at Advanced Photon Source (APS) at Argonne National Laboratory, providing detailed machining error from manufacturer and realistic rough surface. For comparison, a nominal geometry is also used for the simulation. Spray characteristics such as Sauter Mean Diameter (SMD), droplet volume and surface area are extracted by post-processing the CFD outputs. It is seen that compared to the nominal geometry, the use of the high resolution real geometry predicts about 11% lower SMD. The rough surface along with manufacturing defects generate more unstable structures along the jet surface and accelerates the liquid deformation and breakup processes, starting inside of the counter-bore. This result shows that the machining details of injector, which is usually ignored in the two-phase flow simulations, has considerable impact on the spray development process.

## Introduction

For Spark-Ignition (SI) engines, Gasoline direct injection (GDI) system offers a number of advantages over Port Fuel Injection (PFI), such as accurate fuel delivery, less Cycle-to-Cycle Variation (CCV), better fuel economy with extended knock limit operating condition, and potential for stratified lean combustion [1]. The fuel injection process in GDI engine becomes critical to the subsequent fuel/air mixing, spark ignition and flame propagation, and requires better understanding to improve the engine stability, efficiency and emissions.

Compared to diesel injectors, the multi-hole gasoline direct injectors usually feature smaller length/diameter (L/D) ratio, narrower included spray angle, a step-hole design, and operate at a relatively lower fuel pressure (10-40 MPa). The spray atomization and direction are influenced by the in-nozzle flow substantially. Despite much advance in experimental measurement techniques, Computational Fluid Dynamic (CFD) is a unique tool to reveal the phenomenon of interest inside and near-nozzle dense regions. Saha et al. [2] applied the Eulerian mixture model and Homogeneous Relaxation Model (HRM) to simulate the Engine Combustion Network (ECN) Spray G injector with transient moving needle. Partial hydraulic flip was observed, and the spray targeting angle was also shown to be lower than the drill angle due to the backflow of chamber ambient gas into counter-bore. A one-way coupling approach [3] was further developed to couple the internal nozzle Eulerian flow with discrete Lagrangian spray parcels, and showed accurate prediction in spray penetration. Befrui et al. [4] simulated the same injector using Volume-of-Fluid Large-Eddy-Simulation (VOF-LES) approach to predict the jet primary breakup. This approach resolves liquid blob with refined computational mesh and provides information of droplet size, velocity and spray cone angle, which were then used to initialize Lagrangian spray simulation for model validation.

Numerous studies have shown that the internal nozzle geometry has significant impacts on both internal flow and external spray. Duke et al. [5] studied the spray tomography for three 6-hole GDI nozzles with different hole arrangement, using x-ray radiography measurement. Asymmetric structure of spray plumes varying from solid-cone to hollow-cone or crescent shape were observed, which were considered to depend on nozzle geometry such as hole inclination, needle lift and L/D ratio. Matusik et al. [6] measured the nozzle geometry for 8 nominally duplicate Spray G injectors using x-ray computed tomography (CT) imaging. By performing multiple linear regression analysis with respect to average planar integrated mass (PIM), several key dimensions were identified, as shown in Table 1. There are significant deviations between the manufactured and nominal values, which is expected to influence spray process significantly. However, almost all the CFD effort for GDI simulation to date use a nominal injector geometry model, and are used to compare with experimental data measured from real injectors without consideration of such manufacturing errors.

In this study, a high-fidelity LES simulation of Spray G injector is performed using VOF method. An experimentally measured needle movement profile is applied to accurately capture the start-of-injection transient process. A micron-level high resolution injector geometry model is used to include manufacturing tolerance and defects. Moreover, simulation using a nominal geometry is also performed for comparison.

*Table 1 Injector geometry measurement, averaged over 64 samples from Matusik et al. [6]*

| Geometric feature | ID | Mean [μm] | Std. [μm] | Nominal [μm] |
|---|---|---|---|---|
| Hole diameter | D1 | 175 | 1.14 | 165 |
| Hole length | L1 | 150 | 10.5 | 170 |
| Hole length/diameter ratio | L1/D1 | 0.86 | - | 1.03 |
| Hole inlet corner radius | R1 | 4.93 | 0.88 | 0 |
| Counter-bore diameter | D2 | 394 | 1.29 | 388 |
| Counter-bore length | L2 | 402 | 11.18 | 470 |

**Numerical Methods**

Volume of fluid method [7] is applied to model the two-phase flow in this study using Converge$^{TM}$ code [8] In this method, both fluids are resolved on computational mesh and velocity difference across interphase is neglected. The governing equation reads,

$$\rho \left[\frac{\partial \boldsymbol{u}}{\partial t} + \boldsymbol{u} \cdot \nabla \boldsymbol{u}\right] = -\nabla p + \nabla \cdot [\mu \nabla \boldsymbol{u} + \mu(\nabla \boldsymbol{u})^T] + \sigma \kappa \delta_S \boldsymbol{n} \qquad (1)$$

where $\boldsymbol{u}$ is the velocity field, $\rho$ is the density, $p$ is the pressure, $\mu$ is the viscosity, $\sigma$ is the surface tension, $\kappa$ is the local interface curvature, $\delta_S$ is a Dirac distribution concentrated at interface $S$, and $\boldsymbol{n}$ is a unit vector normal to the interface.

In a cell where both phases exist, the following equations hold,

$$\rho = \rho_g \alpha + \rho_l (1 - \alpha) \qquad (2)$$

$$\mu = \mu_g \alpha + \mu_l (1 - \alpha) \qquad (3)$$

The subscript $g$ and $l$ represent gas phase and liquid phase, respectively. $\alpha$ is the void fraction in a cell. In VOF method, the transport equation of $\alpha$ is solved for two-phase immiscible incompressible flows,

$$\frac{\partial \alpha}{\partial t} + \boldsymbol{u} \cdot \nabla \alpha = 0 \qquad (4)$$

Additionally, the piecewise-linear interface calculation (PLIC) method [9] is applied to construct sharp interface between phases.

LES method is used for the turbulent flow modeling. After applying the spatial filtering operation, the Navier-Stokes equation contains an unclosed non-linear term that must be modeled. Dynamic structure model [10] is employed which solves the transport equation for subgrid scale kinetic energy, $k_{sgs}$, and the subgrid stress is modeled as:

$$\tau_{ij} = C_{ij} k_{sgs} \qquad (5)$$

$$C_{ij} = 2 \frac{L_{ij}}{L_{kk}} \qquad (6)$$

where $L_{ij}$ is the modified Leonard stress term.

The GDI nozzle simulated is ECN Spray G injector #28. A full nozzle geometry model with 1.7 μm resolution is measured by x-ray CT imaging at the Advanced Photon Source (APS) of Argonne National Laboratory. A 135 degree sector domain is used to reduce the computational expense, which includes three holes and a spherical chamber region with a radius of 4.5 mm, as shown in Figure 1 (a). However, due to the memory requirements of the computational solver for the real geometry details, only the parts of sac bottom, hole and counter-bore surfaces of hole #5 are used in simulation as indicated by the red component in Figure 1 (b), in conjunction with nominal geometry for the rest parts. Thus, only the spray emitting from the middle hole is used for analysis and discussion, while the two sprays on the sides serve to provide a realistic boundary condition in terms of air-entrainment etc. for the middle one. An enlarged snapshot for hole # 5 can be found in Figure 1 (c). Other than the nozzle dimensions that deviate from the nominal values as shown in Table 1, the real geometry captures the surface roughness, which is approximated at order of 5 micron. Additionally, manufacturing defects such as the protrusion at hole exit, and void on the counter-bore surface, are also preserved and highlighted by the red circle in Figure 1 (c). These details will be shown later since it impacts the spray morphology considerably.

A moving needle is considered with the measured profile as shown in Figure 2. The grey box highlights the 0.1 ms period that is simulated, which covers the whole transient period of start-of-injection. The minimum needle lift is set to be 5 μm at the start of simulation. Above this minimum gap, the in-nozzle region is initialized with fuel at 298 K, 190 bar, with Dirichlet inlet boundary of 298 K, 190 bar. Below the minimum gap, the sac region is initialized with fuel at 298 K, 3.15 bar. The regions of hole, counter-bore, and chamber are initialized with nitrogen at 298 K, 3.15 bar, with open boundary at chamber outlet. A 2.5 μm boundary embedding is applied to the real geometry part to preserve the realistic surface details, and region embedding with minimum cell size of 5 μm is applied for the internal and near nozzle region. Adaptive-mesh-refinement (AMR) of 5 μm is also enabled in the chamber based on the velocity field. The total cell count is ~9 million at the beginning, then increases to ~35 million as the needle lifts and liquid fuel is injected into chamber. With this setup, the current 0.1 ms simulation takes ~78,000 core hour to finish.

VOF simulation predicts spray atomization and droplet breakup processes, and then the 3D output data is further post-processed to extract statistical results. The 3D data is first resampled on a uniform mesh, then void fraction of 0.5 is used as a threshold to classify each cell containing liquid or gas phases. The 26-connectivity algorithm is applied, and liquid cells that touches each other via faces, edges, or corners are considered as neighbors and belong to the same liquid blob. An isosurface is also constructed on each blob. Then, information such as volume, $V$, and surface area, $A$, can be calculated for each liquid structure.

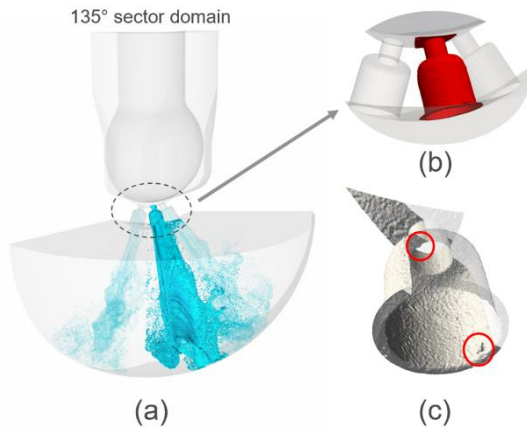

*Figure 1. Geometry model used for CFD simulation.*

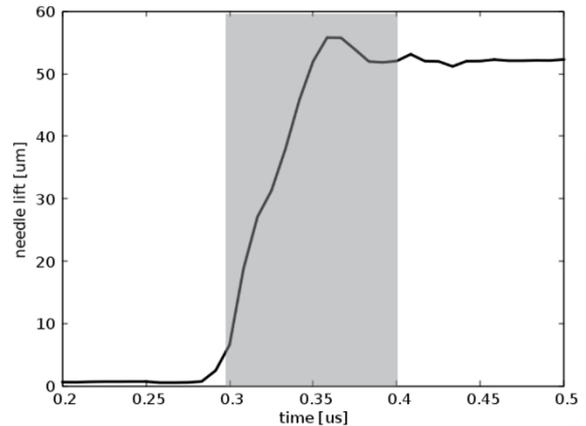

*Figure 2. Needle lift profile vs. time obtained from experimental measurements.*

**Results and Discussion**

Figure 3 shows the 3D visualizations of spray development at 20, 40, 60, 80, and 100 μs after start-of-injection (ASI), respectively. Results of simulation with nominal geometry are shown on the left side, in comparison with results of real geometry on the right. The liquid-gas interface is defined as α=0.5, and shown as the iso-surface in Figure 3. The intact liquid core is coloured by transparent cyan, while any other blob that is not in touch with the liquid core is considered as detached droplet, and is coloured by red.

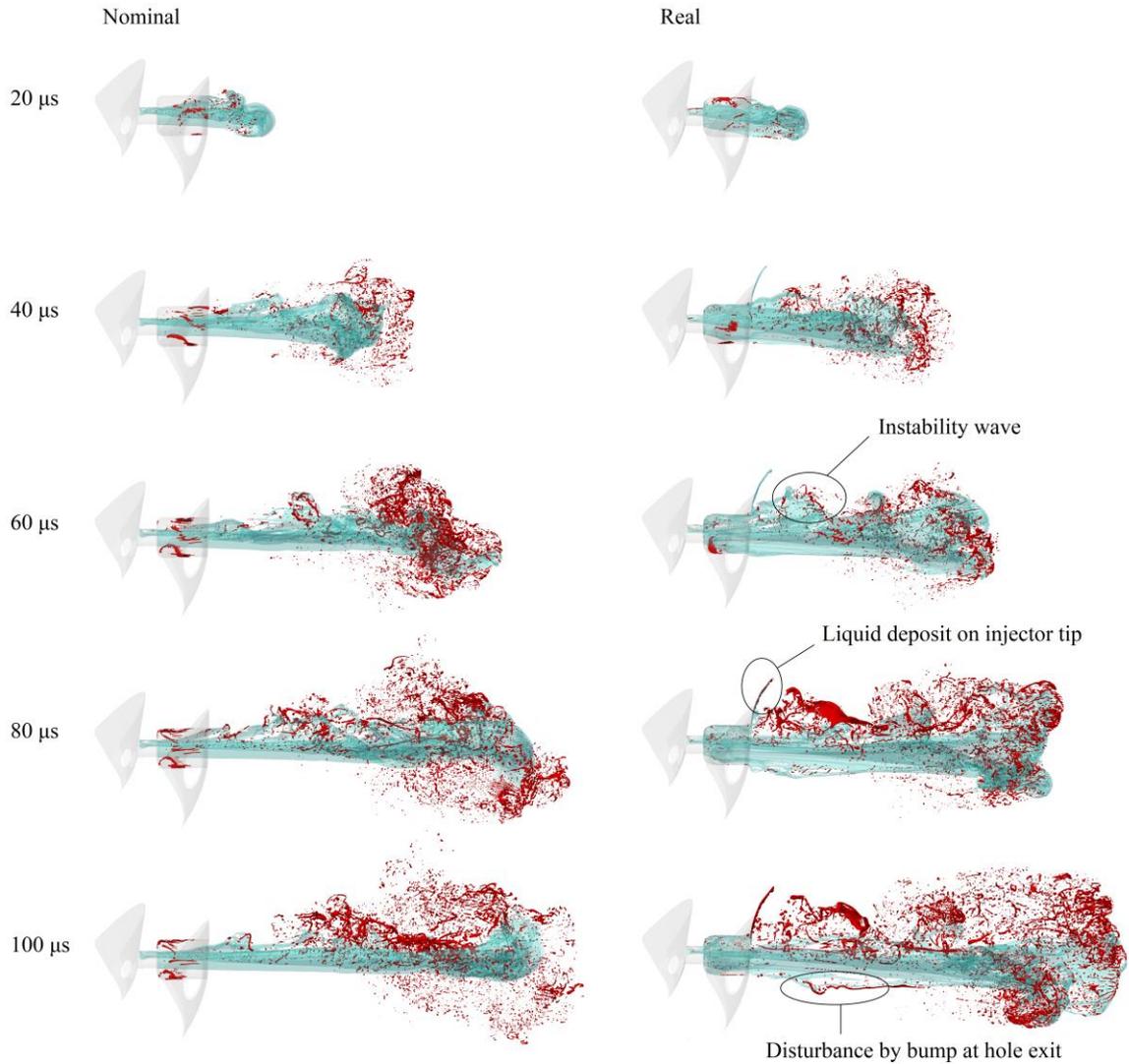

*Figure 3. Visualization of spray evolution at 5 time instants between nominal and real geometries.*

At the early stage of fuel injection event, most of the liquid breakup occurs at the jet head since it experiences a higher level of relative velocity as the jet penetrates into a quiescent environment. Stream-wise ligaments are formed on the head and shed into smaller blobs. Instability waves are seen to develop along the jet surface, mostly on the outer periphery, leading to primary breakup that forms smaller liquid droplets. Interestingly, no unstable structure is found on the inner side of jet with the nominal geometry due to the guidance from smooth hole boundary. However, droplet detachment is seen on the inner side of jet for the real geometry, due to the disturbance from rough surface, especially the bump at hole exit as seen in Figure 1 (c). Overall, the real geometry presents more unstable structures and faster breakup starting inside of the counter-bore. The spray is interacting with the counter-bore wall with considerable amount of liquid deposit on the injector tip, which is not seen with the nominal geometry. It is worth to note that the rate of injection reaches steady value of 2 g/s for the real geometry and 1.9 g/s for the nominal geometry, at 50 µs ASI. The slightly higher rate of injection prediction with real geometry is considered as a result of less sharp hole inlet corner, as well as larger hole diameter as shown in Table 1. The measured rate of injection for standard Spray G is 1.875 g/s [11]. The simulation with the real geometry over-predicts the mass flow rate by 6.7 %, and this is largely because the density (690.1 kg/m$^3$) at cooler fuel temperature of 298 K in simulation is ~9% higher than the density (632.6 kg/m$^3$) at 363 K in experiment.

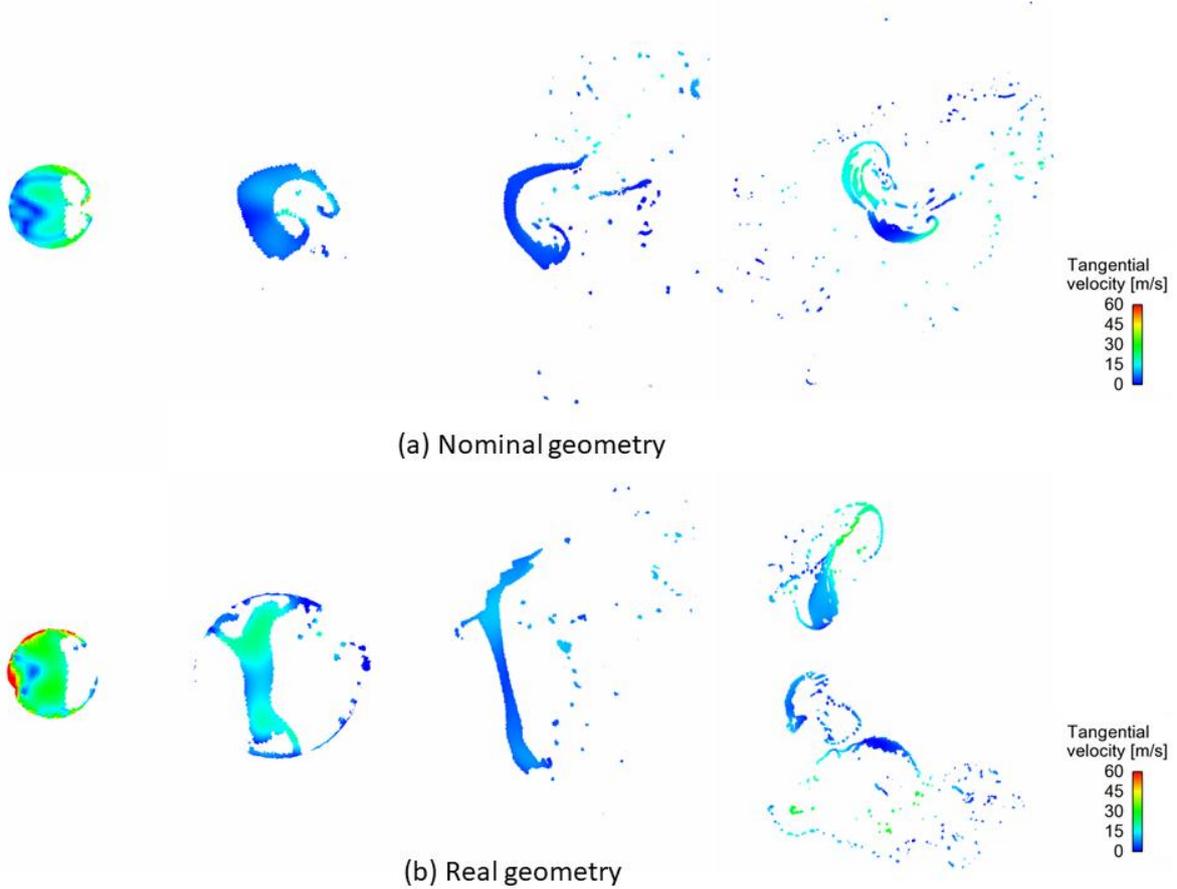

*Figure 4. Liquid spray cross-sections colored by radial velocity magnitude, at (from left to right) hole exit, counter-bore exit, 10 D1 and 20 D1, at t = 100 μs ASI.*

Figure 4 shows the cross-sections of sprays at hole exit, counter-bore exit, 10 D1 and 20 D1 downstream of the orifice exit, which reveals the spray morphology, especially the shape of liquid core. The color contours indicate the magnitude of radial velocity component that is tangential to the cross-section surface. The internal flow separates from the hole wall due to the sharp hole inlet corner, and does not re-attach given that the hole length is short, leaving a small area in the cross-section at hole exit being occupied by the ambient gas. The hydraulic flip leads to liquid core in crescent shape instead of solid cylinder. For the real geometry, the liquid core quickly spreads into a flat sheet, and even bifurcates into two parts in the downstream. The different spray morphology seen between nominal and real geometries is largely due to the significantly different radial velocity at hole exit, especially at the periphery where the liquid is in touch with the boundary. The higher radial velocity originates from the disturbance by the rough surface and leads to higher kinetic energy that is normal to the liquid/gas interface, which evolves into faster instability wave growth, and contributes to highly deformed shape of liquid core and faster primary breakup. Figure 5 shows statistics of the blob count along the axial direction at t = 60 μs ASI. Consistent with the findings in Figure 3 and Figure 4, the spray atomization at upstream is enhanced by the use of the real geometry. Compared to the nominal one, the real geometry predicts a higher blob count by a factor of 2 to 3 at the upstream locations, and a slower penetration rate due to higher momentum loss to the ambient gas with a more dispersed spray.

The 26-connectivity analysis provides information such as volume diameter, $d_v = \sqrt[3]{6V/\pi}$, and surface diameter, $d_s = \sqrt{A/\pi}$. Figure 6 shows the blob size distribution in terms of $d_v$ and $d_s$ for the real geometry, at 20, 60 and 100 μs ASI. A 45° line is drawn to indicate the location where blob is in spherical shape with $d_v = d_s$. It is seen that liquid structures with $d_v < 20\mu m$ closely stick to the 45° line since small liquid droplet tends to remain in spherical shape. Liquid structures with $20 < d_v < 200\mu m$ deviate from the 45° line to the $d_v < d_s$ side, indicating that larger liquid structures are in stretched shape, e.g., ligaments, due to the aerodynamic forces. One liquid structure with $d_v > 200\mu m$ can be found at each time instant, which is considered as the intact liquid core and consists of more than 90% mass of the entire spray. For all time instants, and also for both the real and the nominal geometries, the location in the $d_v$-$d_s$ space for each blob falls into the same trajectory as seen in Figure 6, and $d_v = 200 \mu m$ is found to be a robust threshold to separate the intact liquid core from the detached droplets.

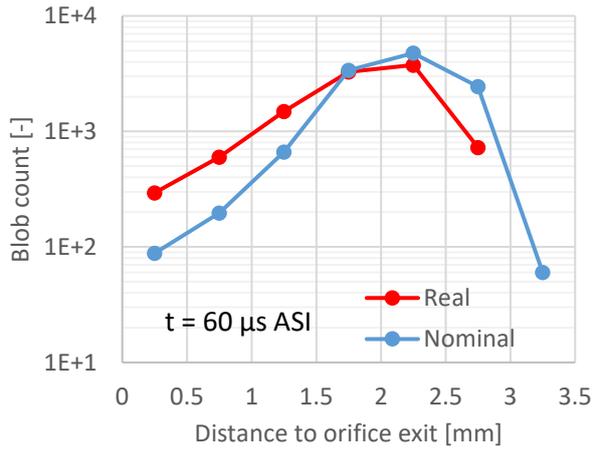

*Figure 5. Blob count in logarithmic scale as function of axial distance at t = 60 μs ASI.*

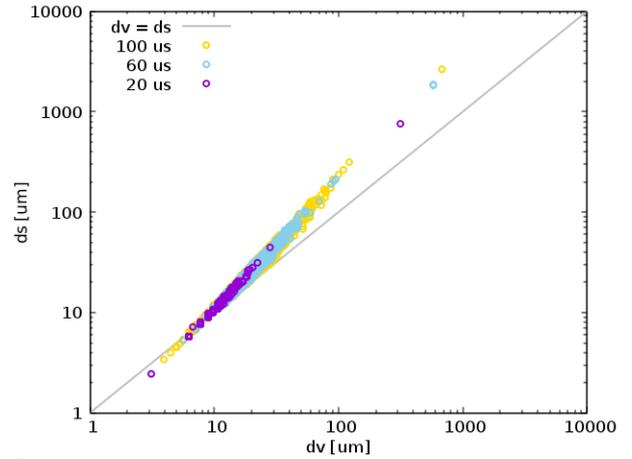

*Figure 6. Blob size distribution in volume diameter - surface diameter space for the real geometry.*

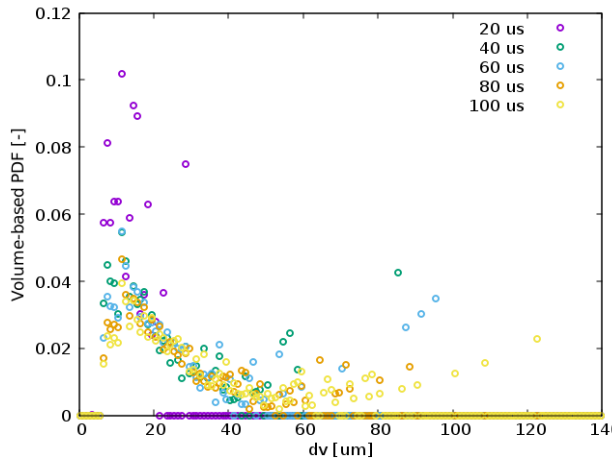

*Figure 7. Volume-based PDF of blob volume diameter for the real geometry.*

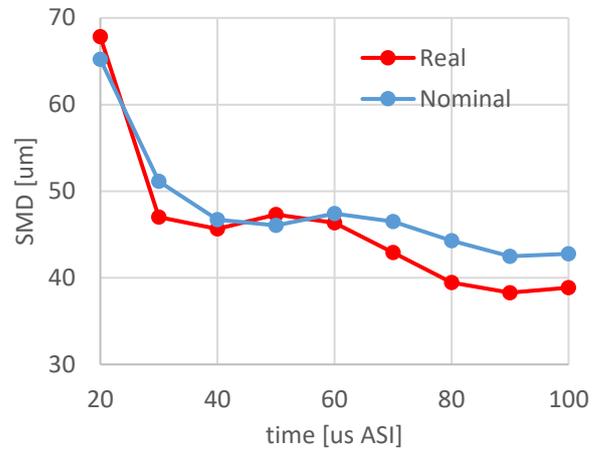

*Figure 8. Overall SMD as function of time between real and nominal geometries.*

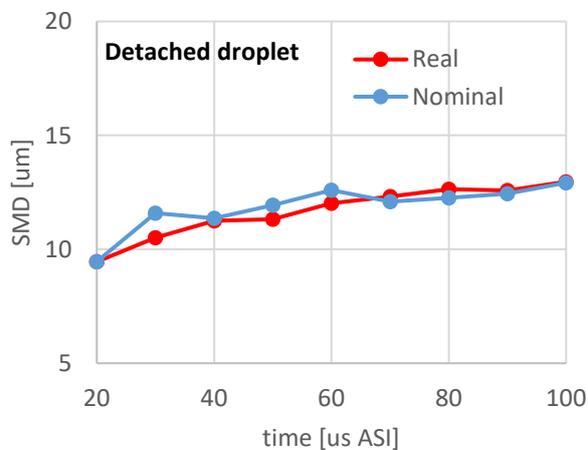

*Figure 9. SMD of detached droplets (dv < 200 μm) for real and nominal geometries.*

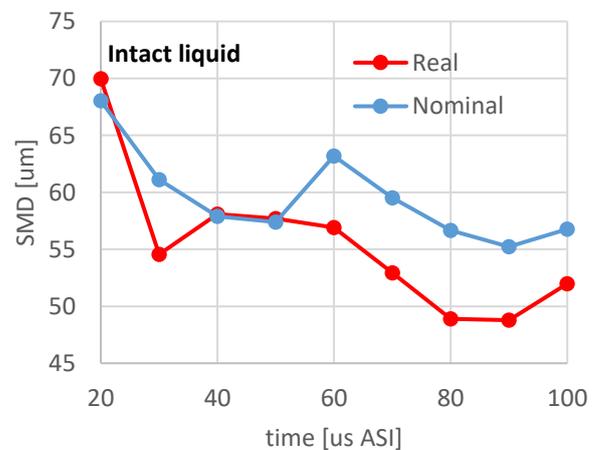

*Figure 10. SMD of intact liquid core (dv > 200 μm) for real and nominal geometries.*

The volume-based probability density function (PDF) of $d_v$ is shown in Figure 7. Except for the very early stage at 20 μs ASI, the PDF profile for blobs with $d_v < 50 \mu m$ is almost unchanged through time, in a form of log-normal distribution with mode at 10-13 μm. Larger blobs are formed at later stage, as results of coalescence under non-vaporizing condition and large liquid structures detach from the intact core, which can be also observed from Figure 3. It is also worth to note that only a limited number of blobs are formed with $d_v > 50 \mu m$, thus the PDF value is zero at some $d_v$ categories as seen in Figure 7, as oppose to the continuous size distribution for the small blobs.

Sauter mean diameter (SMD), defined as $SMD = 6(\sum V_i)/(\sum A_i)$, is plotted as function of time in Figure 8. After a drastic drop from 20 to 30 μs ASI, the SMD decreases slightly from ~50 μm to ~40 μm by the end of simulation. The use of the real geometry results in 11% smaller SMD compared to the nominal geometry. Note the SMD calculation shown in Figure 8 considers all liquid structures including the intact liquid core and the detached droplets. Further breakdown information can be found in Figure 9 and Figure 10, where the SMDs are calculated for the detached droplet ($d_v < 200 \mu m$) and the intact liquid core ($d_v > 200 \mu m$), respectively. The prediction of detached droplet SMD ranges from 10 to 13 μm, which is in good agreement with the droplet phase-doppler measurement at 15 mm downstream of the injector tip [11]. The detached droplet SMD slightly increases over time due to large blobs being formed in later stage, as discussed above. Interestingly, negligible difference in predictions of detached droplet SMD is seen between the real and nominal geometries. However, significant difference is found in intact liquid core SMD, which is the major contributor to the difference in overall SMD seen in Figure 8. As revealed in Figure 4, the shape of liquid core is more stretched with real geometry, leading to higher surface/volume ratio and consequently smaller SMD values.

**Summary and Conclusions**

A high fidelity LES simulation of ECN Spray G injector is presented in this work. The VOF model coupled with the PLIC method is applied to capture the spray atomization and droplet breakup with sharp liquid/gas interface construction. A real injector geometry model is used, which was measured by x-ray CT imaging with micron resolution and provides detailed surface and manufacturing tolerance. For comparison, a nominal geometry is also used in simulation. The predicted rate of injection and detached droplet SMD are in good agreement with the experimental measurement. Compared to the nominal geometry, the real geometry predicts faster primary breakup and liquid core deformation at the upstream region starting inside of the counter-bore, and predicts 11% smaller overall SMD. Such significant difference can be well explained by the radial velocity distribution at spray cross-section. The enhanced radial velocity is directly related to the disturbance by the realistic rough surface and manufacturing defects that are observed with the real geometry. However, there is also study [6] showing that the manufacturing tolerance in several design features is important to the spray development. In order to better understand the critical geometry feature for spray development, numerical study with several realistic geometries is suggested for future work to isolate the surface roughness effect and the manufacturing tolerance.

**Acknowledgements**

UChicago Argonne, LLC, operator of Argonne National Laboratory ("Argonne"), a U.S. Department of Energy Office of Science laboratory, is operated under Contract No. DE-AC02- 06CH11357. The U.S. Government retains for itself, and others acting on its behalf, a paid-up nonexclusive, irrevocable worldwide license in said article to reproduce, prepare derivative works, distribute copies to the public, and perform publicly and display publicly, by or on behalf of the Government. This research was partially funded by DOEs Office of Vehicle Technologies, Office of Energy Efficiency and Renewable Energy under Contract No. DE-AC02-06CH11357. The authors wish to thank Gurpreet Singh, and Kevin Stork, program managers at DOE, for their support. This research was conducted as part of the Co-Optimization of Fuels & Engines (Co-Optima) project sponsored by the U.S. Department of Energy (DOE) Office of Energy Efficiency and Renewable Energy (EERE), Bioenergy Technologies and Vehicle Technologies Offices.

The authors would like to thank Dr. Katarzyna Matusik and Dr. Christopher Powell from APS of Argonne for providing the high-resolution injector geometry model.